\documentclass{appolb}

\usepackage{bm} 
\usepackage{graphicx}

\preprint{}

\begin{document}

\title{Resonance production in a thermal model\thanks{Talk presented at the
XXXIII International Symposium on Multiparticle Dynamics,
5-11~September~2003, Cracow, Poland} 
\thanks{Supported in part by the Polish State Committee for 
Scientific Research, grant 2~P03B~059~25.}}
\author{\underline{W. Broniowski}$^a$ and W. Florkowski$^{a,b}$
\address{$^a$The H. Niewodnicza\'nski Institute of Nuclear Physics,\\
Polish Academy of Sciences, PL-31342 Cracow, Poland}
\address{$^b$Institute of Physics, \'Swi\c{e}tokrzyska Academy, PL-25406 Kielce, Poland}}

\maketitle

\begin{abstract}
We discuss the $\pi^+ \pi^-$ invariant-mass correlations and the 
resonance production in a thermal model with single freeze-out. 
The predictions are confronted with the recent data from the STAR Collaboration.
\end{abstract}

\thispagestyle{empty}

Detection of hadronic resonances at RHIC (see the contribution of Fachini and 
Refs.~\cite{starkstar,starrho}) provides important
clues on the particle production mechanism and subsequent evolution of the system formed
at mid-rapidity in relativistic heavy-ion collisions. 
In this talk we analyze the $\pi^+\pi^-$ invariant-mass correlations in the
framework of the single-freeze-out model of Ref.~\cite{sf}. For more details 
the reader is referred to Ref.~\cite{resonance}.

The thermal or statistical approach to heavy-ion physics has been 
very instructive and successful in classifying and understanding the data
({\em cf.} contributions of Calderon, Ga\'zdzicki, Hama, Kisiel, Ster, and Becattini). 
We recall briefly the basic assumptions of our model:
\begin{enumerate} 
\item An approximation of a single freeze-out, $T_{\rm chem}=T_{\rm kin} \equiv T$, is made.  
This has recently gathered experimental support, with the short lifetime
of the hadronic phase obtained from the HBT measurements of $R_{\rm out}/R_{\rm side}\sim 1$ 
and from the 
out-of-plane elongation of $R_{\rm side}$ ({\em cf.} contributions of Appelsh\"auser and Lisa).
\item A {\em complete} treatment of resonances, with all particles from the Particle Data Tables
incorporated.
\item The shape of the freeze-out hypersurface is assumed in the 
spirit of the Buda-Lund model \cite{BL}. It possesses a Hubble-like flow, 
with the four-velocity proportional to the 
coordinate, {$u^\mu=x^\mu/\tau$}.
\item The model has altogether four parameters. The thermal ones, {$T=160$~MeV and $\mu_B=26$~MeV} at
$\sqrt{s_{NN}}=200$~GeV, 
are fixed by fitting the ratios of the particle abundances. The geometry
parameters, namely the 
invariant time at freeze-out {$\tau$}, which controls the overall normalization, 
and transverse size {$\rho_{\rm max}$} are fixed by fitting the $p_\perp$ spectra.
\end{enumerate}

In order to analyze the $\pi^+\pi^-$ invariant mass spectra measured in 
Ref.~\cite{starrho} with the like-sign subtraction technique, we make the assumption that 
all the correlated pion pairs are obtained from the decays of neutral resonances: 
$\rho$, $K_S^0$, $\omega$, $\eta$, $\eta'$, $f_0/\sigma$, and $f_2$. 
The phase-shift formula for the volume density of resonances with spin degeneracy
$g$ is given by the formula \cite{phsh} 
\begin{equation}
\frac{dn}{dM} = g \int \frac{d^3 p}{(2\pi)^3} \frac{d {\delta_{\pi \pi}(M)}}{\pi dM} 
\frac{1}{\exp \left( \frac{\sqrt{M^2+{\bm p}^2}}{T} \right ) \pm 1},
\end{equation}
and was used in Ref.~\cite{resonance}. We note that the same formalism has been also applied by
Pratt and Bauer \cite{Pratt} in a similar analysis.
In some works the spectral function of the resonance is used as the
weight, instead of the derivative of the phase
shift. For narrow resonances this does not make a difference, since
then {$d\delta(M)/dM \simeq \pi \delta(M-m_R)$}.
For wide resonances, such as in the scalar-isoscalar channel, 
or for effects of tails, the difference
between the correct formula and the one with the spectral function is significant.

In order to get a feeling on the problem, we first carry on a warm-up calculation,
where a static source is used. 
Fig.~1 shows the mid-rapidity invariant-mass spectra computed at two
different values of the freeze-out temperature: $T=165$~MeV and $110$~MeV. 
We note a clear appearance 
of the included resonances, as well as a feature of a 
much steeper fall-off of the strength in Fig.~1(b), which simply 
reflects the lower value of the temperature.
\begin{figure}
\vspace{-9mm}
\begin{center}
\includegraphics[width=9.7cm]{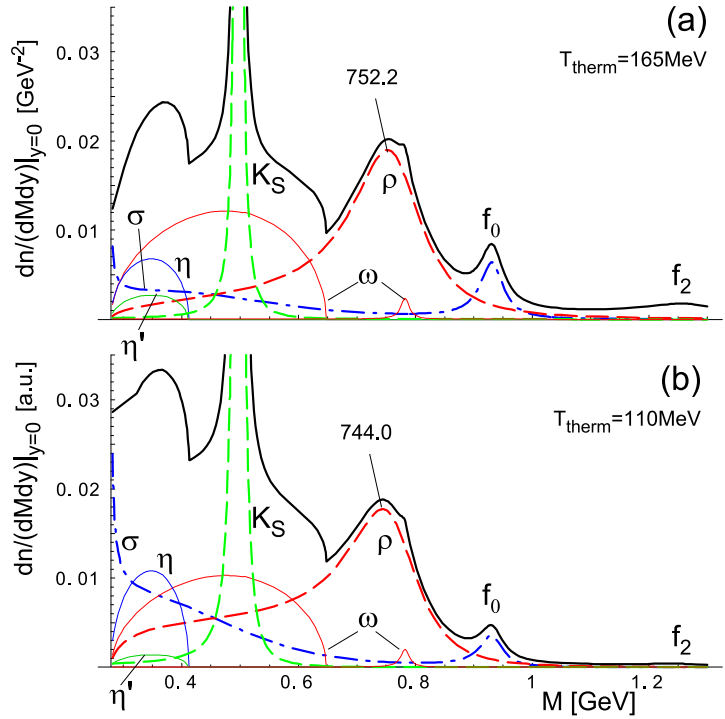}
\end{center}
\vspace{-5mm}
\caption{Calculation of the $\pi^+ \pi^-$ invariant-mass correlation from decays of 
resonances in a thermal model with a static source. (a) The case
  of $T_{\rm therm}=T_{\rm chem}=165~{\rm MeV}$.  (b) The case of
  $T_{\rm therm}=110~{\rm MeV}$. The numbers indicate the position of
  the $\rho$ peak (in MeV) and the labels indicate the decay channel.}
\end{figure}
Thus, the $\pi^+ \pi^-$ invariant-mass spectrum provides yet another {\em thermometer}
for the decoupling temperature.

The collective flow has no effect on the
invariant mass of a pair of particles produced in a resonance decay,
since the quantity is Lorentz-invariant. Nevertheless, it affects the
measured results since the kinematic cuts in an obvious manner break this invariance. 
We have performed a full-fledged calculation in our model, including the 
flow, kinematic cuts, and decays of higher resonances.
The results are shown in Ref.~\cite{resonance}. One of the thrilling 
aspects of the analysis is whether the observed shift of the $\rho$ peak 
can be explained. We have found that the thermal effects and the kinematic cuts 
are able to shift it down by about 30~MeV, leaving a few tens of MeV unexplained.
For that reason, we have repeated the calculation with the position of the $\rho$
peak moved down by 9\%.  
In Fig.~2 we compare the model predictions to the STAR data.  
Our results were scaled by a common factor, which is equivalent to 
fitting the value of $\tau$ for the peripheric collisions of the STAR experiment. 
Then they were filtered by the detector efficiency 
correction (we are grateful here to Patricia Fachini). We note that the model 
does a very good job in reproducing the gross features of the data. We also note that
the full model, with feeding from higher resonances and with flow/cuts, which
has $T=165$~MeV, gives similar predictions
to the naive model at $T=110$~MeV rather that 165~MeV. 
\begin{figure}
\vspace{-9mm}
\begin{center}
\includegraphics[width=10.5cm]{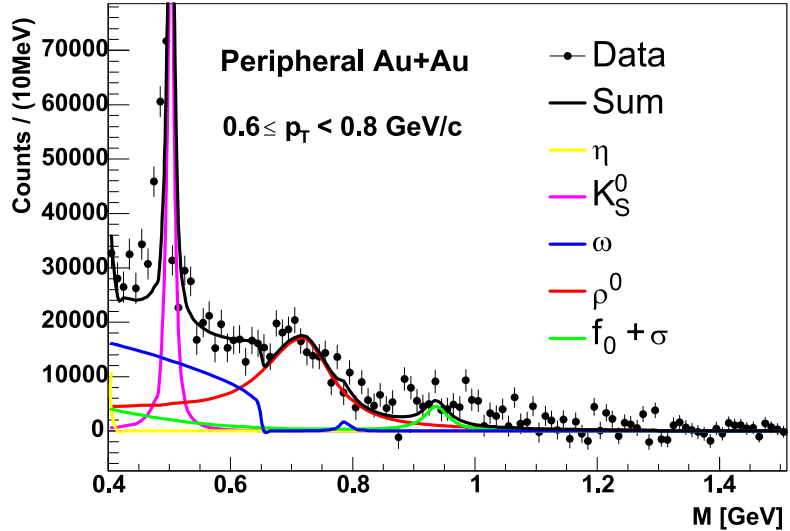}
\end{center}
\vspace{-9mm}
\caption{Single-freeze-out model vs. the data of Ref.~\cite{starrho} for the $\pi^+ \pi^-$ 
invariant-mass correlations 
(from {\tt http://www.star.bnl.gov/$\sim$pfachini/KrakowModel}). The model calculation includes the 
decays of higher resonances, the flow, the kinematic cuts, and the detector efficiency.}
\end{figure}
This is yet another manifestation of the effect of {\em ``cooling''} induced by the resonances
\cite{michal}.

More details and further results and discussion concerning the 
abundances and the transverse-momentum spectra 
of various resonances can be found in Ref.~\cite{resonance}.

\end{document}